\documentstyle[preprint,aps]{revtex}

\begin{document}
\title{
\hfill
\parbox{4cm}{\normalsize KUNS-1288\\HE(TH)~94/12 \\hep-lat/9409011}\\
\vspace{1.0cm}
Boundary condition for Staggered Fermion \\
in Lattice Schr\"odinger Functional of QCD
}

\author{ \normalsize \sc
Shunji Miyazaki\thanks{e-mail address:
miyazaki@gauge.scphys.kyoto-u.ac.jp}
\
\hskip 6pt and  \hskip 6pt
Yoshio Kikukawa\thanks{e-mail address:
kikukawa@gauge.scphys.kyoto-u.ac.jp}
}

\address{\normalsize \em
         Department of Physics, Kyoto University\\
         Kyoto 606-01, Japan
}

\date{\normalsize September, 1994}

\maketitle

\vskip 36pt
\begin{abstract}
\baselineskip 20pt
The fermionic part of the Schr\"odinger functional of QCD
is formulated in the lattice regularization with the staggered fermion.
The boundary condition imposed on the staggered fermion field
is examined in terms of the four-component Dirac spinor.
The boundary terms are different from those of the Symanzik's theory
in the flavor structure due to the species doubling.
It is argued that,
in the case of the homogeneous Dirichlet boundary condition,
surface divergence does not occur
if the link variables of gauge field are introduced on the
original lattice, not on the blocked one.
Its application to the numerical calculation of the running coupling
constant in QCD is discussed.
\end{abstract}

\vskip 16pt
\newpage

\newcommand{\bra}[1]{\left\langle #1\right\vert}
\newcommand{\ket}[1]{\left\vert #1\right\rangle}
\newcommand{\sch}{Schr\"odinger Functional}
\newcommand{\st}{staggered fermion}
\newcommand{\gam}{\gamma}
\newcommand{\am}{\Lambda^4}
\newcommand{\x}{{\bbox{x}}}
\newcommand{\y}{{\bbox{y}}}
\newcommand{\bn}{{\bbox{n}}}
\newcommand{\brho}{{\bbox{\rho}}}
\newcommand{\pbrho}{{\bbox{\rho^\prime}}}
\newcommand{\bx} {{\bbox{x}}}
\newcommand{\4}{ {n_4} }
\newcommand{\half}{\frac{1}{2}}
\newcommand{\quart}{\frac{1}{4}}
\newcommand{\ee}{\, .}
\newcommand{\ec}{\, ,}
\newcommand{\Tr}{ {\rm Tr}}
\newcommand{\R}[2]{(R_{\rho}^{})_{#1 \ }^{\ #2}}
\newcommand{\Rd}[2]{(R_{\rho}^\dagger)_{#2 \ }^{\ #1}}
\newcommand{\pR}[2]{(R_{\rho^\prime}^{})_{#1 \ }^{\ #2}}
\newcommand{\pRd}[2]{(R_{\rho^\prime}^\dagger)_{#2 \ }^{\ #1}}
\newcommand{\sR}{R_{\rho^\prime}^{}}
\newcommand{\sRd}{R_{\rho^\prime}^\dagger}
\newcommand{\delt}[2]{\delta_{#1}^{#2}}
\newcommand{\sitarel}[2]{ {\mathrel{\mathop{\kern0pt #1}\limits_{#2}}} }
\newcommand{\uerel}[2]{ {\mathrel{\mathop{\kern0pt #1}\limits^{#2}}} }

\subsection*{\normalsize \bf Introduction}

Numerical calculation of the strong coupling constant $\alpha_s$ in QCD
has been recently attempted by several groups.
These works involve various physical and technical ingredients,
namely, charmonium spectrum\cite{el-khadra}, static quark-antiquark
force\cite{michael}\nocite{schilling}--\cite{bali}
and the Schr\"odinger functional of SU(3)
Yang-Mills theory in finite volume\cite{luscher}\nocite{su2}--\cite{su3}.
To fix the scale,
low energy physical quantities were measured
in the unit of the lattice spacing
or
in the unit of the linear extent of the finite volume.
The scale was related to
the running coupling constant in the
(modified) minimal-subtraction scheme
by the improved lattice perturbation theory\cite{mack}
or
through the nonperturbatively defined running coupling constant.
The basic Monte Carlo simulations were all
performed in the quenched approximation (the pure Yang-Mills theory).
As to the charmonium spectrum, however,
the potential model has allowed one to estimate systematic errors for
the approximation\cite{quark-potential}.

The desired next step is surely the calculation of $\alpha_s$
based on the simulation with dynamical fermion.
In this respect,
the sea quark effect on the charmonium system
has been observed by Onogi et al. \cite{onogi}\cite{seaquark}
and their result is consistent with the procedure
adopted by Elkhadra~et~al..
Quite recently, a new determination
using the $\Upsilon$ spectrum and
including the effect of dynamical quarks
was reported by Davies~et~al.\cite{upsilon}.
These works actually suggest that the precise determination
of $\alpha_s$ through the quarkonium spectrum is promising.

On the other hand, the finite size scaling technique adopted in the works
of L\"uscher~et~al. can be also applied to QCD.
The Schr\"odinger functional of QCD in
the lattice regularization has been recently formulated by Sint\cite{sint}.
He examined the boundary condition of fermion field
in a finite extent of the time direction
for the case of the Wilson fermion\cite{wilson}.
He derived the boundary terms and
argued that the resulted boundary terms are not lattice artifacts
and meaningful at the continuum limit.
He also examined the spectrum of the Dirac operator of
the free fermion and its squared.
He has observed that the boundary condition forces a minimal frequency
in the system and gives a chance to simulate QCD with light quarks.

Generally speaking,
there exists the arbitrariness of choice for the boundary condition
of the fermion field. Actually, in his original paper
on the Schr\"odinger functional\cite{symanzik},
Symanzik adopted a different
boundary condition from that is derived from the Wilson fermion.
Concerning the lattice regularization,
another type of lattice fermion, the staggered
fermion\cite{susskind}, is known to work.
Then a different boundary condition is expected for it.

As to the staggered fermion, however,
the four-fold degeneracy of flavor may cause us
both good and harm in calculating $\alpha_s$.
It may be useful and economical in the simulation for the
ideal case of four degenerate massless quarks.
However, as a final goal, we need to put the realistic mass
to each quark:
a few MeV for the up quark on one side, but about one GeV
for the charm quark on the other side.
The possibility of giving the non-degenerate mass term
to the staggered fermion has been discussed by several
authors\cite{mitra}\nocite{mitra:nucl}\nocite{mitra-wise}--\cite{gockeler}.
But it is yet an open question to give such large
mass-difference to the staggered fermion in actual numerical
computations.
More technically, it may need a different treatment
for the staggered fermion
of the boundary effect due to the finite extent of the space-time.

In this paper,
we will formulate
the fermionic part of
the Schr\"odinger Functional of QCD with the staggered fermion
through its transfer matrix,
which is formulated by Thun~et~al.~\cite{thun},
and work out its boundary condition.
Then we will discuss the possibility of getting
along with the staggered fermion in the calculation
of $\alpha_s$ in QCD.

\subsection*{\normalsize \bf Transfer Matrix for the Staggered Fermion}

The Schr\"odinger Functional in lattice QCD
can be naturally formulated by the transfer matrix as shown by Sint for
the case of the Wilson fermion\cite{sint}.
The transfer matrix for the staggered fermion has been given
by H.~S.~Sharatchandra, H.~J.~Thun and P.~Weisz\cite{thun}.
Firstly we will review their results and fix the notation that we use.
The lattice spacing $a$ is chosen for unity, whereas
the lattice sites are labeled by four integers
$n_{\mu}=(n_1,n_2,n_3,n_4)=(\bn ,n_4)$.
$\hat{\mu}$ stands for a unit vector of the $\mu $-th direction.
Since the mass terms are found not to be important when one constructs
the transfer matrix of the staggered fermion,
they are omitted here.
Then the action of free massless staggered fermion is given by
\begin{equation}
S_F=\frac{1}{2}\sum_{n, \mu} \eta_\mu(n) \, \bar{\chi}(n)
[ \chi (n+\hat{\mu})- \chi (n-\hat{\mu})]
\ec
\label{staggered:free-action:massless}
\end{equation}
where $\chi(n)$ and $\bar\chi(n)$ are one-component Grassmann
variables\cite{chodos}\cite{kawamoto}\cite{thun}.
$\eta_\mu(n)$ is defined by
\begin{equation}
\eta_\mu (n)=(-1)^{n_1+\cdots +n_{\mu -1}} \ec \quad  \eta_1=1 \ee
\end{equation}
It is related to the Euclidean Dirac gamma matrices satisfying
\begin{equation}
\{ \gamma^{\mu},~\gamma^{\nu} \}=2\delta_{\mu \nu} \ec
\qquad \gamma^{\mu \dag}=\gamma^{\mu} \ec
\end{equation}
by the equation
\begin{equation}
T(n)\gamma_\mu T^\dagger(n+\hat{\mu})=\eta_\mu (n)
\, \mbox{\boldmath $I$} \ec
\end{equation}
where
\begin{equation}
T(n)=\gamma_{1}^{n_1}\gamma_2^{n_2}\gamma_3^{n_3}\gamma_4^{n_4} \ee
\end{equation}

First of all, we rescale and denote the
variables $\chi(n)$ and $\bar\chi(n)$
such as
\begin{equation}
\chi_{n_4}^{ }(\bn) \equiv \chi(n) \, , \qquad\qquad
\chi_{n_4}^\dagger(\bn) \equiv  \frac{1}{2} \, \bar{\chi}(n)\eta_4(n) \ee
\label{staggered:field-rescale}
\end{equation}
Since $\eta_{\mu}(n)$ does not depend on $n_4$, we can define
\begin{equation}
\eta_{k}'(\mbox{\boldmath $n$})\equiv \eta_{k}(n)\eta_4(n) \ec
\end{equation}
and it follows that
\begin{equation}
\eta_{k}'(\bn + \hat{\mbox{\boldmath $k$}}) = -\eta_{k}'(\bn) \ee
\end{equation}
Then the action (\ref{staggered:free-action:massless}) is rewritten as
\begin{eqnarray}
S_F
&=&
\sum_{\4,\bn}
\{ \chi_{n_4}^\dagger(\bn) \chi_{n_4+1}^{ }(\bn)
  +\chi_{n_4}^{ }(\bn)     \chi_{n_4+1}^\dagger(\bn)  \}
\nonumber \\
&+&
\sum_{\4,\bn}\sum_{k}
\{ \chi_{n_4}^\dagger(\bn) \eta_{k}'(\bn)
   \chi_{n_4}^{ }(\bn+\hat{\mbox{\boldmath $k$}})
+\chi_{n_4}^\dagger(\bn+\hat{\mbox{\boldmath $k$}})
\eta_{k}'(\mbox{\boldmath$n$})\chi_{n_4}^{ }(\bn)     \}
\label{staggered:free-action:massless:scaled}
\end{eqnarray}

The next step is the characteristic one for the staggered fermion:
we introduce a set of auxiliary fields
$ \varphi$ and $ \varphi^\dagger$
using the $\delta$ function for Grassmann variable.
Namely, we use\footnote{
We have adopted the convention for the measure
of the Grassmann integral as
$\int\!d\varphi^\dagger d\varphi \; \varphi \varphi^\dagger = 1 $
and the abbreviation,
$\left[ d\varphi_{n_4}^\dagger d\varphi_{n_4}^{ }\right]\equiv \prod_{\bn}
\big\{ d\varphi_{n_4}^\dagger (\bn)d\varphi_{n_4}^{ }(\bn) \big\}$.
}
\begin{equation}
\mbox{\boldmath $I$}=
\int  \prod_\4
\left[ d\varphi_{n_4}^\dagger d\varphi_{n_4}^{ } \right]
\prod_{\bn,\4}
\delta(\varphi_\4^{ }(\bn)-\chi^\dagger_\4(\bn))
\delta(\varphi^\dagger_\4 (\bn)-\chi_\4^{ }(\bn))
\ec
\end{equation}
where
\begin{eqnarray}
\delta(\varphi^\dagger_\4 (\bn)-\chi_\4^{ }(\bn))
&\equiv& (\varphi^\dagger_\4 (\bn)-\chi_\4^{ } (\bn))
\ee
\end{eqnarray}
Then the first two terms in the action
(\ref{staggered:free-action:massless:scaled})
can be written as
\begin{equation}
\sum_{n_4}\sum_\bn
\{ \chi_{n_4}^\dagger(\bn)\chi_{n_4+1}^{ }(\bn)
+\varphi_{n_4}^\dagger(\bn)\varphi_{n_4+1}^{ }(\bn) \} \, .
\label{staggered:free-action:massless:scaled:kinetic-term}
\end{equation}

Accordingly, we introduce two independent sets of operators,
$\hat\chi$, $\hat\chi^\dagger$ and $\hat\varphi$, $\hat\varphi^\dagger$
satisfying the anti-commutation relations as
\begin{equation}
\{ \hat{\chi}(\bn) \, , \, \hat{\chi}^\dagger(\bn') \}=\delta_{\bn,\bn'}
\ec \qquad
\{ \hat{\varphi}(\bn) \, , \,
                        \hat{\varphi}^\dagger(\bn') \}=\delta_{\bn,\bn'}
\ec
\end{equation}
and otherwise zero.
On the Fock space spanned by these operators, we can consider
the coherent states given by
\begin{eqnarray}
\ket{ \chi_{\4+1}^{ },\varphi_{\4+1}^{ } } &\equiv&
\exp \sum_{\bn} \left\{
\hat\chi^\dagger(\bn) \chi_{\4+1}(\bn)
+\hat\varphi^\dagger(\bn) \varphi_{\4+1}(\bn) \right\} \ket{0} \ec \\
\bra{ \chi^\dagger_\4,\varphi^\dagger_\4 } &\equiv & \bra{0}
\exp \sum_{\bn} \left\{
\chi^\dagger_\4(\bn) \hat\chi(\bn)
+\varphi^\dagger_\4(\bn) \hat\varphi(\bn) \right\} \ee
\end{eqnarray}
It can be shown that these coherent states satisfy the following
completeness relation.
\begin{eqnarray}
\mbox{{\boldmath  $I$}}
&=&\int
\left[ d\chi_\4^\dagger d\chi_{\4+1} \right]
\left[ d\varphi_\4^\dagger d\varphi_{\4+1} \right] \nonumber\\
&\times & \exp \left\{ -\sum_{\bn}
[\chi_{n_4}^\dagger(\bn)\chi_{n_4+1}^{ }(\bn)
+\varphi_{n_4}^\dagger(\bn) \varphi_{n_4+1}^{ }(\bn)] \right\}
\ket{\chi_{n_4+1}^{ },\varphi_{n_4+1}^{ } }
\bra{ \chi_{n_4}^\dagger,\varphi_{n_4}^\dagger} \ee
\label{staggerd:coherent-state-completeness}
\end{eqnarray}

We can see that the weight appearing in
Eq.~(\ref{staggerd:coherent-state-completeness})
can be identified with the kinetic term in the $\4$-direction
in the form of
Eq.~(\ref{staggered:free-action:massless:scaled:kinetic-term}).
The remaining terms in the exponential of the action
can be arranged into the product of
the representatives of the transfer matrices on the coherent state basis.
The transfer matrix acting on the Fock space is
then found to be
\begin{eqnarray}
\widehat{T}_F
&=&
\exp \left\{ -\frac{1}{2} \sum_{\bn}\sum_{k=1}^{3}
\Big[ \hat{\chi}^\dagger(\bn+\hat{\mbox{\boldmath $k$}})
      \eta_{k}'(\bn)
      \hat{\varphi}^\dagger(\bn)
     +\hat{\chi}^\dagger(\bn)
      \eta_{k}'(\bn)
      \hat{\varphi}^\dagger(\bn+\hat{\mbox{\boldmath $k$}}) \Big]
    \right\}
\nonumber \\
&& \times
\prod_{\bn}
:\left( \hat{\varphi}^\dagger(\bn)-\hat{\chi}(\bn) \right)
 \left( \hat{\varphi}(\bn)-\hat{\chi}^\dagger(\mbox{\boldmath$n$})
                                                            \right):
\nonumber \\
&& \ \times
\exp \left\{ -\frac{1}{2} \sum_{\bn}\sum_{k=1}^{3}
\Big[ \hat{\varphi}(\bn+\hat{\mbox{\boldmath $k$}})
      \eta_{k}'(\bn)
      \hat{\chi}(\bn)
     +\hat{\varphi}(\bn)
      \eta_{k}'(\bn)
      \hat{\chi}(\bn+\hat{\mbox{\boldmath $k$}}) \Big] \right\}
\ee
\end{eqnarray}
$\widehat{T}_F$ is hermitian, but not positive definite.
The Hamiltonian follows $(\widehat{T}_F)^2$ and it is positive definite.

\subsection*{\normalsize \bf Schr\"odinger Functional}

Hereafter we will consider the staggered fermion
in a three-dimensional Euclidean lattice with finite volume $(2l)^3$
and its time development in a finite time interval $2m$.
The entire four dimensional Euclidean lattice can be represented by
\begin{equation}
\Gamma=
\{
\, n_\mu = (\bn,n_4) \, | \ 0\leq n_k < 2l \, (k=1,2,3),\ 1 \leq n_4 \leq 2m
\, \} \ee
\end{equation}
We assume that the fermionic variables $\chi$, $\chi^\dagger$
and $\varphi$, $\varphi^\dagger$ obey spatially periodic boundary
condition. Temporarily, certain configurations in terms of Grassmann
variables are assumed on initial and final equitime surfaces.
Since the Dirac equation is of first order,
the boundary condition should fix
only half of the components of the field variable
on each initial or final equitime surface.
The choice of the half components
is intimately related to the way of
constructing the Schr\"odinger Functional.

The wave functional in the Schr\"odinger representation
is naturally introduced as the representative of a state vector
in the coherent-state basis,
\begin{equation}
\Psi[\chi^\dagger,\phi^\dagger] \equiv
\left\langle\chi^\dagger,\varphi^\dagger \Big\vert \Psi \right\rangle \ee
\end{equation}
Its dynamics is described by the kernel
which acts on an initial wave functional at $\4=1$
and results in a final wave functional at $\4=2m$.
For the staggered fermion,  it is given by
\begin{equation}
{\cal Z}_F [\chi_{2m}^\dagger,\varphi_{2m}^\dagger;\chi_1,\varphi_1]
\equiv
\bra{ \chi^\dagger_{2m},\varphi^\dagger_{2m}}
(\widehat{T}_F)^{2m}
\Big\vert \chi_{1},\varphi_{1} \Big\rangle
\ee
\end{equation}
It is this kernel which is referred to
the Schr\"odinger functional in the literature.

The Schr\"odinger functional
can be represented by functional integral as follows.
Inserting the completeness relation
(\ref{staggerd:coherent-state-completeness}) in-between the
transfer matrices,
and integrating over the auxiliary fields
$\varphi_{\4+1}^{ }$, $\varphi_{n_4}^\dagger (\4=1,\cdots,2m-1)$  and
$\chi_1^\dagger, \chi_{2m}$ to eliminate the $\delta $-functions,
we obtain
\begin{eqnarray}
{\cal Z}_F[\chi_{2m}^\dagger,\varphi_{2m}^\dagger;\chi_1,\varphi_1]
&=&
\int  \prod_{\4=1}^{2m-1}
\left[ d\chi_\4^\dagger d\chi_{\4+1}^{ }
       d\varphi_\4^\dagger d\varphi_{\4+1}^{ } \right]
\prod_{\4=1}^{2m}
\bra{ \chi_\4^\dagger ,\varphi_\4^\dagger } \widehat{T}_F
\ket{ \chi_\4^{ },\varphi_\4^{ } }
\nonumber\\
&&\qquad\quad
\times
\exp
\biggr\{
-\sum_{\bn,\4=1}^{2m-1}
[\chi_\4^\dagger(\bn)\chi_{\4+1}^{ }(\bn)
+\varphi_\4^\dagger(\bn)
\varphi_{\4+1}^{ }(\bn)]
\biggl\}
\nonumber\\
&=&
\int  \prod_{\4=2}^{2m-1}
\left[ d\chi_\4^\dagger d\chi_\4^{ } \right]
\exp \big\{ -S_{F}^{tot} \big\} \ee
\label{staggered:schrodinger-functional:free}
\end{eqnarray}
$S_{F}^{tot}$ consists of three parts,
\begin{equation}
S_{F}^{tot}
\equiv S_{B}^{(T)}+S_F+S_{B}^{(0)} \ee
\label{staggered:schrodinger-functional:total-action}
\end{equation}
If we use the original conjugate variables
$\chi(\bn,\4)$ for $1\le \4 \le 2m-1$ and
$\bar\chi(\bn,\4)$ for $2\le \4 \le 2m$,
which are rescaled back from
$\chi_{n_4}^{ }(\bn)$ and $\chi_{n_4}^\dagger(\bn)$
by Eq.~(\ref{staggered:field-rescale}),
and we also redefine
\begin{eqnarray}
\bar{\chi}_{1}(\bn) =  2 \eta_4(n) \varphi_{1}(\bn) , \qquad\quad
\chi_{2m}(\bn) = \varphi_{2m}^\dagger(\bn) \ec
\end{eqnarray}
then, these three parts read
\begin{eqnarray}
S_{F}
&=&\sum_{n_4=2}^{2m-1}\sum_{\bn}\sum_{\mu =1}^{4}
\half \eta_\mu (n) \, \bar{\chi}(n)
[ \chi (n+\hat{\mu})- \chi (n-\hat{\mu})] \ec
\label{staggered:schrodinger-functional:action}
\\
S_B^{(T)}
&=&\sum_{\bn}\sum_{k=1}^{3}\half
\eta_{k}(\bn, 2m) \,
 \bar{\chi}(\bn,2m)
[ \chi(\bn+\hat{\mbox{\boldmath $k$}},2m)
- \chi(\bn-\hat{\mbox{\boldmath $k$}},2m)] \nonumber \\
&&{}-\sum_{\bn}\half \eta_{4}(n) \bar{\chi}(\bn,2m) \chi(\bn,2m-1) \ec
\label{staggered:schrodinger-functional:boundary:final}
\\
S_B^{(0)}
&=&\sum_{\bn}\sum_{k=1}^{3}\half
\eta_{k}(\bn,1) \,
 \bar{\chi}(\bn,1)
[  \chi(\bn+\hat{\mbox{\boldmath $k$}},1)
 - \chi(\bn-\hat{\mbox{\boldmath $k$}},1)] \nonumber \\
&&{}+\sum_{\bn}\half  \eta_{4}(n) \bar{\chi}(\bn,1)\chi(\bn,2) \ee
\label{staggered:schrodinger-functional:boundary:initial}
\end{eqnarray}

\subsection*{\normalsize \bf
Boundary Condition in terms of Four-component Spinor}

Starting from the transfer matrix of the staggered fermion,
we have obtained the functional integral form of
the \sch\ for the free quarks.
The first part of the action, $S_F$, given by
Eq.~(\ref{staggered:schrodinger-functional:action}),
is the usual action for the staggered fermion (in finite volume).
$S_B^{(0)}$ and $S_B^{(T)}$ are boundary terms involving
the boundary values of the fermion field, $\chi_1,\bar\chi_1$
and $\chi_{2m}, \bar\chi_{2m}$, respectively.
In order to examine these boundary terms more closely and
to communicate with the counterparts in the continuum theory,
we will express them in terms of the four-flavored four-component
Dirac spinor\cite{gliozzi}\cite{kluberg}.
We follow here the formulation given by
H.~Kluberg-stern~et~al.\cite{kluberg}.

The original index of the lattice $n_{\mu}$ may be written in the form of
\begin{equation}
n_{\mu}=2x_{\mu}+\rho_{\mu} \ec
\end{equation}
where $x_{\mu}$ is an integer four-vector and
$\rho_{\mu}$ is a four-vector whose components are
either one or zero.
$x_{\mu}$ labels the site on the sublattice
with the spacing twice as large as the original one.
$\rho_{\mu}$ points one of the sixteen variables
$ \chi (2x+\rho)$ near to $x_{\mu}$ making them associated
to the site on the sublattice.
Accordingly, we adopt the notation such as
\begin{equation}
\chi_{\rho}(x)=\chi (2x+\rho ) ,\quad
\bar{\chi}_{\rho}(x)=\bar{\chi}(2x+\rho ) .
\end{equation}
It follows immediately that
\begin{eqnarray}
\chi (2x+\rho +\hat\mu)&=&\sum_{\rho '}[\chi_{\rho '}(x)
\delta_{\rho ',\rho +\hat\mu}+\chi_{\rho '}(x+\hat\mu)
\delta_{\rho ',\rho -\hat\mu}] \, , \\
\chi (2x+\rho -\hat\mu)&=&\sum_{\rho '}[\chi_{\rho '}(x)
\delta_{\rho ',\rho -\hat\mu}+\chi_{\rho '}(x-\hat\mu)
\delta_{\rho ',\rho +\hat\mu}] \, ,
\end{eqnarray}
Then, noting that
$\eta_\mu(2x+\rho )=\eta_\mu(\rho )$,
we can rewrite the action $S_F$ as
\begin{eqnarray}
&&
S_F=\frac{1}{4} \sum_{x_4=1}^{m-1}\sum_{\x}\sum_{\rho\rho^\prime\mu}
\left\{ \bar\chi_{\rho}(x)
(\Lambda_{\mu})_{\rho \rho'}
\left[ \chi_{\rho^\prime}(x+\hat\mu) - \chi_{\rho^\prime}(x-\hat\mu)
\right] \right. \\
&&\qquad\qquad\qquad\qquad +
\left. \bar{\chi}_{\rho}(x)
(\Lambda_{\mu}^{5})_{\rho \rho '}
\left[ \chi_{\rho^\prime}(x+\hat\mu) + \chi_{\rho^\prime}(x-\hat\mu)
- 2\chi_{\rho^\prime}(x) \right] \right\}
\ec
\end{eqnarray}
where
\begin{eqnarray}
( \Lambda_{\mu} )_{\rho \rho '}&=&\eta_{\mu}(\rho )
[\delta_{\rho -\hat{\mu},\rho '}+\delta_{\rho +\hat{\mu},\rho '}] \ec
\label{Lambda:def}
\\
( \Lambda^5_\mu )_{\rho \rho '}&=&\eta_{\mu}(\rho )
[\delta_{\rho -\hat{\mu},\rho '}-\delta_{\rho +\hat{\mu},\rho '}] \ee
\label{Lambda5:def}
\end{eqnarray}

These $\Lambda_\mu$'s can be expressed by the traces over the
products of gamma matrices.
First we introduce matrices
\begin{equation}
R_{\rho}^{}= (R_{\rho}^{})_{\alpha \ }^{\ a}
\equiv \half \, T(2x+\rho ) =
\half \,
\gamma_{1}^{\rho_1}\gamma_2^{\rho_2}\gamma_3^{\rho_3}\gamma_4^{\rho_4}
\ec
\end{equation}
satisfying unitary conditions
\begin{eqnarray}
\sum_{\alpha a} \Rd{\alpha}{a}
                \pR{\alpha}{a}
&=&
\Tr \left\{ R_\rho^\dagger R_{\rho^\prime}^{} \right\}
=
\delta_{\rho \rho^\prime}
\ec
\nonumber \\
\sum_{\rho} \R{\alpha}{a}
            \Rd{\beta}{b}
&=&\delt{\alpha}{\beta} \delt{b}{a}
\ee
\end{eqnarray}
Then we can show the following relations.
\begin{eqnarray}
(\Lambda_\mu)_{\rho \rho^\prime}
&=&
\Tr
\left\{
R_\rho^\dagger \gamma_\mu R_{\rho^\prime}^{}
\right\} \ec
\label{Lambda:gamma}\\
(\Lambda^5_\mu)_{\rho \rho '}
&=&
\Tr
\left\{
R_\rho^\dagger \gamma_5 R_{\rho^\prime}^{} \gamma_5 \gamma_\mu
\right\}
\ec
\label{Lambda5:gamma}\\
\sum_{\rho\rho^\prime}
\R{\alpha}{a}
(\Lambda_\mu)_{\rho \rho^\prime}
\pRd{\beta}{b}
&=&
(\gamma_\mu)_\alpha^{\ \beta} \otimes \bbox{1}^{\ a}_b
\quad\quad \equiv (\Gamma_\mu)_{\alpha \ b}^{\ \beta \ a}
\ec
\label{Lambda:tensor}\\
\sum_{\rho\rho^\prime}
\R{\alpha}{a}
(\Lambda^5_\mu)_{\rho \rho^\prime}
\pRd{\beta}{b}
&=&
(\gamma_5)_{\alpha}^{\ \beta}
\otimes
(\gamma_\mu^\ast \gamma_5^\ast)^{\ a}_{b}
\equiv (\Gamma^5_\mu)_{\alpha \ b}^{\ \beta \ a}
\label{Lambda5:tensor}\ee
\end{eqnarray}

With this unitary matrices $R_\rho^{}$, we can transform
$\chi_\rho(x)$ into the four-flavored Dirac spinor as
\begin{eqnarray}
\psi_\alpha^{\ a}(x)
&=&
\frac{1}{ \sqrt{2}}\sum_{\rho}
\R{\alpha}{a} \chi_{\rho}(x)
\ec
\nonumber \\
\bar{\psi}_a^{\ \alpha}(x)
&=&
\frac{1}{  \sqrt{2}}\sum_{\rho}
\bar{\chi}_{\rho}(x)
\Rd{\alpha}{a}
\label{four-spinor}
\ec
\end{eqnarray}
where Greek and Latin suffices can be taken to
denote spinor and flavor indices, respectively.
In terms of the four-component spinor thus defined, we obtain\footnote{
The normalization in Eq.~(\ref{four-spinor}) is determined
so as to scale all dimension-full quantities in the four
component spinor representation by the new lattice
spacing $b (=2a)$ twice as large as $a$.}
\begin{eqnarray}
S_F
&=&\sum_{x_4=1}^{m-1}\sum_{\x \, \mu}
 \bar{\psi}(x)
\left[
(\gamma_{\mu}\otimes \mbox{\bf 1})
\widetilde\nabla_{\mu}
+\half
(\gamma_5 \otimes \gamma_\mu^\ast \gamma_5^\ast )
 \bigtriangleup_\mu
\right] \psi(x) \ec
\end{eqnarray}
where
\begin{eqnarray}
\widetilde\nabla_\mu \psi (x)
&\equiv &
\half [\psi (x+\hat{\mu})-\psi(x-\hat{\mu})] \ec
\\
\bigtriangleup_\mu \psi(x)
&\equiv&
\psi(x+\hat{\mu})+\psi(x-\hat{\mu})-2\psi(x) \ee
\end{eqnarray}

Now we consider the surface term, $S_{B}^{(T)}$,
at $x_4=m\equiv T/b$.
It can be rewritten in the form of
\begin{eqnarray}
S_{B}^{(T)}&=&-\sum_{\x}\sum_{\brho \pbrho}
\quart \; \bar{\chi}^{ }_{(\brho,0)}(\x,m) \,
( \Lambda_4-\Lambda^5_4 )_{(\brho,0)(\pbrho,1)} \,
\chi^{ }_{(\pbrho,1)}(\x,m-1)
\nonumber \\
&&{}+
\sum_{\x}\sum_{\brho \pbrho}\sum_{k=1}^{3}
\half
\bar{\chi}^{ }_{(\brho,0)}(\x,m)
\left[
 (\Lambda_k)_{(\brho,0)(\pbrho,1)}
\widetilde\nabla_{k}
+\half  (\Lambda^5_k)_{(\brho,0)(\pbrho,1)}
 \bigtriangleup_k
\right]
\chi^{ }_{(\brho,0)}(\x,m)
\ee
\label{staggered:schrodinger-functional:boundary:four-spinor0:final}
\end{eqnarray}
Here we denoted the four-vector $\rho$ by a set of
a three-vector and a definite forth component as
$(\brho,0)$ or $(\brho,1)$.
It is useful to introduce
projectors which act on the space of the four-vectors \{$\rho$\}
and project those elements with a definite value for the forth
component $\rho_4$.
We define
\begin{equation}
(\widetilde P_0)_{\rho \rho^\prime}
\equiv\delta_{\rho+\hat{4},\rho^\prime+\hat{4}} \ec
\qquad
(\widetilde P_1)_{\rho \rho^\prime}
\equiv\delta_{\rho-\hat{4},\rho^\prime-\hat{4}}
\ee
\end{equation}
The relations below follow the definitions of
Eqs. (\ref{Lambda:def}) and (\ref{Lambda5:def}):
\begin{eqnarray}
\sum_{\sigma}
(\Lambda_\mu)_{\rho \sigma } (\Lambda_\mu)_{\sigma \rho^\prime}
&=&\delta_{\rho+\hat{\mu} \rho^\prime+\hat{\mu}}
+\delta_{\rho-\hat{\mu} \rho^\prime -\hat{\mu}}
=\delta_{\rho,\rho^\prime}
\qquad (\mu=1,\cdots, 4)
\ec
\\
\sum_{\sigma}
(\Lambda_\mu)_{\rho \sigma } (\Lambda^5_\mu)_{\sigma \rho^\prime}
&=&\delta_{\rho+\hat{\mu} \rho^\prime+\hat{\mu}}
-\delta_{\rho-\hat{\mu},\rho '-\hat{\mu}}
\quad\qquad\qquad (\mu=1,\cdots, 4)
\ee
\end{eqnarray}
By using these relations, the above projectors can be expressed
by $\Lambda_4$ and $\Lambda^5_4$
as follows.
\begin{eqnarray}
(\widetilde P_0)_{\rho \rho^\prime}
&=&\half \sum_{\sigma}
(\Lambda_4)_{\rho \sigma }
(\Lambda_4 + \Lambda^5_4)_{\sigma \rho^\prime}
\ec\\
(\widetilde P_1)_{\rho \rho^\prime}
&=&\half \sum_{\sigma}(\Lambda_4)_{\rho \sigma }
( \Lambda_4-\Lambda^5_4)_{\sigma \rho^\prime}
\ee
\end{eqnarray}
Making use of these projectors,
Eq.~(\ref{staggered:schrodinger-functional:boundary:four-spinor0:final})
can be written as
\begin{eqnarray}
S_{B}^{(T)}&=&
- \quart \sum_{\x}\sum_{\rho,\rho^\prime}
\bar{\chi}_\rho (\bx,m)
\left[
\widetilde P_0
( \Lambda_4-\Lambda^5_4)
\widetilde P_1
\right]_{\rho\rho^\prime}
\chi_{\rho^\prime}(\bx,m-1)
\nonumber \\
&&{} +
 \half \sum_{\x}
\sum_{\rho \rho^\prime}
\sum_{k=1}^{3}
\bar{\chi}_{\rho}(\bx,m)
\left[
\widetilde P_0
(\Lambda_k \widetilde\nabla_k+\half \Lambda^5_k
\bigtriangleup_k)
\widetilde P_0
\right]_{\rho\rho^\prime}
\chi_{\rho^\prime} (\bx,m)
\ee
\end{eqnarray}

With Eqs.~(\ref{Lambda:tensor}), (\ref{Lambda5:tensor})
and (\ref{four-spinor}),  we can again rewrite it in
terms of the four-component spinor.
\begin{eqnarray}
S_{B}^{(T)}
&=&
- \sum_{\x}
\bar\psi(\x,m) P_0
\Gamma_4 \, \psi(\x,m-1)
\nonumber \\
&&{} +
\sum_{\x} \sum_{k=1}^{3}
\bar{\psi}(\x,m) P_0
(\Gamma_k \widetilde\nabla_k
+\half \Gamma^5_k \bigtriangleup_k ) \,
\psi(\x,m)
\ec
\label{staggered:schrodinger-functional:boundary:four-spinor:final}
\end{eqnarray}
where $P_0$ and $P_1$ are the projectors
acting on the spinor and flavor spaces,
which correspond to $\widetilde P_0$ and $\widetilde P_1$, respectively:
the unitary transformation by $R$ leads to the following expressions.
\begin{eqnarray}
P_0&=&\half \Gamma_4 (\Gamma_4 +\Gamma^5_4)=\half (
\mbox{\bf 1}\otimes\mbox{\bf 1}
+\gamma_4 \gamma_5 \otimes\gamma_4^\ast \gamma_5^\ast )~,
\label{projector:zero}\\
P_1&=&\half \Gamma_4 (\Gamma_4 -\Gamma^5_4)=\half (
\mbox{\bf 1}\otimes\mbox{\bf 1}
-\gamma_4 \gamma_5 \otimes\gamma_4^\ast \gamma_5^\ast )
\label{projector:one}\ee
\end{eqnarray}
Here we have used the anti-commutation relations of $\Gamma$.
\begin{eqnarray}
\{ \Gamma_\mu\, ,\, \Gamma_\nu \}&=&\ \ 2\delta_{\mu \nu}
\,   \mbox{\bf 1}\otimes\mbox{\bf 1}  \ec  \\
\{ \Gamma^5_\mu\, ,\, \Gamma^5_\nu \}&=&-2\delta_{\mu \nu}
\, \mbox{\bf 1}\otimes\mbox{\bf 1}  \ec \\
\{ \Gamma_\mu\, ,\, \Gamma^5_\nu \}&=&\ \ 0
\ee
\end{eqnarray}
By a similar consideration, $S_{B}^{(0)}$ can be
expressed in the form of
\begin{eqnarray}
S_{B}^{(0)}&=& \sum_{\x}
\bar{\psi}(\x,0) P_1 \Gamma_4 \psi(\x,1)
\nonumber \\
&+&
\sum_{\x}\sum_{k=1}^{3}
\bar{\psi}(\x,0) P_1
( \Gamma_k \widetilde\nabla_k
+\half \Gamma^5_k \bigtriangleup_k) \psi(\x,0)
\ee
\end{eqnarray}
Now we summarize the total action
(\ref{staggered:schrodinger-functional:total-action})
in terms of the four-flavored four-component spinor:
\begin{eqnarray}
S_{F}^{tot}
&=&
- \sum_{\x}
\bar{\psi}(\x,m) P_0 \Gamma_4 \psi(\x,m-1)
\nonumber \\
&&\quad +
\sum_{\x}\sum_{k=1}^{3}
\bar{\psi}(\x,m) P_0
(  \Gamma_k \widetilde \nabla_k
+\half
\Gamma^5_k \bigtriangleup_k) \psi(\x,m)
\nonumber \\
&&+
\sum_{x_4=1}^{m-1}\sum_{\x}
\sum_{\mu =1}^{4}
\bar{\psi}(x)
[ \Gamma_\mu \widetilde\nabla_\mu
+\half \Gamma^5_\mu \bigtriangleup_\mu ]
\psi(x)
\nonumber \\
&&+
\sum_{\x}
\bar{\psi}(\x,0)
P_1 \Gamma_4 \psi(\x,1)
\nonumber \\
&&\quad +
\sum_{\x} \sum_{k=1}^{3}
\bar{\psi}(\x,0)
P_1 (  \Gamma_k \widetilde\nabla_k
+\half  \Gamma^5_k \bigtriangleup_k)
\psi(\x,0)
\ee
\label{staggered:schrodinger-functional:total-action:four-spinor}
\end{eqnarray}
The field components to be fixed by
the boundary conditions,
$\chi_1$, $\bar{\chi}_1$ at $x^4=0$ and
$\chi_{2m}$, $\bar{\chi}_{2m}$ at $x^4=T/b$,
can be expressed by the projectors acting on the
four-flavored Dirac spinor as
\begin{eqnarray}
P_1 \psi(\x,0)
&=&\rho_0 (\x) \ec
\quad
\bar{\psi}(\x,0) P_1 =\bar{\rho}_0(\x) \, ,
\nonumber
\\
P_0 \psi(\x,m)
&=&\rho_T (\x) \, ,
\quad
\bar{\psi}(\x,m) P_0 =\bar{\rho}_T(\x) \, ,
\label{staggered:four-spinor:boundary:inhomogeneous}
\end{eqnarray}
where $\rho_t(\x)$ and $\bar\rho_t(\x)$ $(t=0,T)$
are certain Grassmann valued functions.

In the classical continuum limit,
the coordinate $x_\mu$ and the field $\psi(x)$ are
scaled by the lattice spacing $b$ as
\begin{equation}
y_\mu = b x_\mu \ec
\qquad \Psi(y)= \frac{1}{b^{3/2}} \psi(x) \ec
\qquad \eta(\y) = \frac{1}{b^{3/2}} \rho(\x) \ee
\end{equation}
Note also that, in the lattice formulation,
it is the component
$P_0 \psi(x)$ at $x_4=+1$
that has the coupling to $\bar\rho_0(\x)$,
but the same component $P_0 \psi(x)$ at $x_4=0$ does not exist.
In the classical continuum limit $b \rightarrow 0$,
the former comes close to $\bar\rho_0(\x)$ to have a local coupling.
We will express this situation with '$+0$',
namely,
$P_0 \Psi(\y,+0)
=\sitarel{\lim}{b\rightarrow0}\frac{1}{b^{3/2}}P_0\psi(\x,+1)$.
The same is true for the component $P_1 \Psi(y)$ at $x_4=T$.
Then the action
(\ref{staggered:schrodinger-functional:total-action:four-spinor})
of the staggered fermion
reads in the classical continuum limit
\begin{eqnarray}
S_{F}^{tot}&=&\int_{0}^{T}dy_4 \int_{0}^{L}d^3y
\bar{\Psi}(y)
\left[
  (\gamma_\mu \otimes \mbox{\bf 1} ) \partial_{\mu}
\right]
\Psi(y) \, \bigg\vert_{\rm H.B.C.}
\nonumber \\
&-&
\int_{0}^{L}d^3y
\left\{
  \bar{\Psi}(\y,+0)  \Gamma_4 P_1 \, \eta^{ }_0(\y)
- \bar\eta^{ }_0(\y) \Gamma_4 P_0 \, \Psi(\y,+0)
\right\}
\nonumber \\
&-&
\int_{0}^{L}d^3y
\left\{
 \bar\eta^{ }_T(\y) \Gamma_4 P_1 \, \Psi(\y,T-0)
-\bar\Psi(\y,T-0)   \Gamma_4 P_0 \, \eta^{ }_T(\y)
\right\}
\ec
\label{staggered:schrodinger-functional:total-action:four-spinor:continuum}
\end{eqnarray}
where H.B.C. stands for the homogeneous boundary condition with
$\eta(\y)$ and $\bar\eta(\y)$ vanishing:
\begin{eqnarray}
P_1 \Psi(\y,0) &=&0  \, ,
\quad
\bar{\Psi}(\y,0)P_1 =0 \, .
\nonumber
\\
P_0 \Psi(\y,T) &=&0  \, ,
\quad
\bar{\Psi}(\y,T) P_0 = 0 \, .
\label{staggered:four-spinor:continuum:boundary:homogeneous}
\end{eqnarray}
It is also clear
in the lattice formulation that,
in the view point of the path integral,
the integral variables are $\Psi(y)$ and $\bar\Psi(y)$
supplemented with the homogeneous boundary condition
(\ref{staggered:four-spinor:continuum:boundary:homogeneous}), and
$\eta(\y)$ and $\bar\eta(\y)$ which are localized on the boundary surfaces
can be regarded as external sources.

A few remarks are in order here.
It is interesting to note that
the boundary condition for the staggered fermion
(\ref{staggered:four-spinor:boundary:inhomogeneous}) and
the projectors
(\ref{projector:zero}) and (\ref{projector:one}),
are very similar to those Symanzik adopted for
fermion field
in his paper on the Schr\"odinger representation of the
renormalizable field theory\cite{symanzik}.
According to Symanzik's prescription,
the boundary terms of the fermion field in a renormalizable
theory can be constructed in the Minkowski space as follows.
We first find out the operator which
generates the time reversal transformation for fermion.
Such an operator may be chosen as
\begin{equation}
B
=  i \int d^3y \,
\bar \Psi(y) \gamma_0 (- i \gamma_0 \gamma_5) \Psi(y)
=  i \int d^3y \,
\left\{
\bar \Psi(y) \gamma_0  P_- \Psi(y)
- \bar \Psi(y) \gamma_0  P_+ \Psi(y)
\right\}
\ec
\end{equation}
where $P_\pm \equiv (1\pm i \gamma_0 \gamma_5) /2 $,
and the transformation is given by
\begin{equation}
i \left[ B \, , \, P_- \Psi(y) \right] = + P_- \Psi(y)
\ec
\quad
i \left[ B \, , \, P_+ \Psi(y) \right] = - P_+ \Psi(y)
\ee
\end{equation}
The component without a change of the sign is referred to
``Dirichlet component''
and the component with a change of the sign is
``Neumann component''.
First consider imposing the homogeneous Dirichlet boundary condition;
\begin{equation}
P_- \Psi(\y,+0) = 0 \ec\quad \bar \Psi(\y,+0) P_-  =0
\end{equation}
at $y_0=0$. (Here the region $y_0 > 0$
is referred to ``Dirichlet side''. )
It can be achieved by adding the following boundary term
to the action as an interaction.
\begin{equation}
B_0
=  i \int d^3y \,
\left\{
 \bar \Psi(\y,+0) \gamma_0  P_- \Psi(\y,-0)
-\bar \Psi(\y,-0) \gamma_0  P_+ \Psi(\y,+0)
\right\}
\ee
\label{symmanzik:surface-counterterm}
\end{equation}
To implement
the inhomogeneous Dirichlet boundary condition,
\begin{equation}
      P_- \Psi(\y,+0) = \eta(\y)         \ec\quad
\bar \Psi(\y,+0) P_-  = \bar\eta(\y)     \ec
\end{equation}
the Dirichlet component valued functions,
$\eta(\y)$ and $\bar\eta(\y)$,
are introduced as the sources which are localized
at the boundary surface and
the Neumann components in the Dirichlet side are made to couple
with the sources.
Namely, the following source term is added to the action:
\begin{equation}
B_\eta
=  i\int d^3y \,
\left\{
 \bar \Psi(\y,+0) \gamma_0  P_- \eta(\y)
-\bar \eta(\y) \gamma_0  P_+ \Psi(\y,+0)
\right\}
\ee
\end{equation}

We can now clearly see the correspondence between the
boundary condition for the staggered fermion field and
the one that is imposed on the continuum fermion field
in the Symanzik's theory:
the time reversal transformation can be associated
to the matrix
$\pm\left(\gamma_4\gamma_5 \otimes\gamma_4^\ast\gamma_5^\ast \right)$,
and accordingly, the projectors
$P_{0,1}=(1 \pm \gamma_4\gamma_5 \otimes\gamma_4^\ast\gamma_5^\ast)/2$
enter the boundary terms.
Except for the structure in the flavor space
due to the species doubling,
the boundary condition is essentially the same.

\subsection*{\normalsize \bf Gauge Interaction}

The gauge field part
of the Schr\"odinger Functional of QCD can be formulated just as
that of the pure Yang-Mills theory given by
L\"uscher~et~al.\cite{luscher} and
that of QCD with the Wilson fermion given by Sint\cite{sint}.
In the case of the staggered fermion $\chi(n)$,
one can introduce the gauge link variables on the lattice
of the spacing $a$.
In this case, there is no essential difference in constructing
the transfer matrix and the Schr\"odinger Functional
from the case of the free staggered fermion\cite{thun}.

It is also possible, following Kluberg-stern et al.\cite{kluberg},
to define the four-component Dirac spinor in a gauge invariant manner.
Although the action 
in terms of the four-component Dirac spinor
can be obtained only in the expansion with respect to $a$,
this is enough in order to read off the boundary condition
for the fermion field in the classical continuum limit.
The gauge invariant four-spinor can be defined by
\begin{equation}
\psi_\alpha^{\ a}(x)
=
\frac{1}{ \sqrt{2}}\sum_{\rho}
\R{\alpha}{a} U_\rho(x) \chi_{\rho}(x)
\ec
\end{equation}
where
\begin{equation}
U_\rho(x) = U_1(2x)^{\rho_1}
            U_2(2x+\rho_1 \hat1)^{\rho_2}
            U_3(2x+\rho_1\hat1+\rho_2\hat2)^{\rho_3}
            U_4(2x+\rho_1\hat1+\rho_2\hat2+\rho_3\hat3)^{\rho_4}
\ee
\end{equation}
The action reads in the classical continuum limit
\begin{eqnarray}
S_{F}^{tot}&=&\int_{0}^{T}dy_4 \int_{0}^{L}d^3y
\bar{\Psi}(y)
\left[
  (\gamma_\mu \otimes \mbox{\bf 1} ) D_{\mu}
\right]
\Psi(y) \, \bigg\vert_{\rm H.B.C.}
\nonumber \\
&-&
\int_{0}^{L}d^3y
\left\{
  \bar{\Psi}(\y,+0) \Gamma_4 P_1  \eta^{ }_0(\y)
- \bar\eta^{ }_0(\y) \Gamma_4 P_0 \Psi(\y,+0)
\right\}
\nonumber \\
&-&
\int_{0}^{L}d^3y
\left\{
 \bar\eta^{ }_T(\y) \Gamma_4 P_1 \Psi(\y,T-0)
-\bar\Psi(\y,T-0) \Gamma_4 P_0 \eta^{ }_T(\y)
\right\}
\ec
\label{staggered:schrodinger-func:total-action:four-spinor:continuum:gauge}
\end{eqnarray}
where
\begin{equation}
D_{\mu} \equiv \partial_{\mu}+A_{\mu}(y) .
\end{equation}
The boundary condition is the same as Eqs.
(\ref{staggered:four-spinor:continuum:boundary:homogeneous}).

\subsection*{\normalsize \bf Boundary counterterms}

We will next discuss
the possible appearance of divergence at the surfaces by interaction.
According to Symanzik\cite{symanzik},
since the effect of the boundary surface can be expressed by
the local surface interaction as
Eq.~(\ref{symmanzik:surface-counterterm}),
it is possible to apply the ordinary prescription for
the (perturbative) renormalization with the local counterterms,
as far as the theory concerning is renormalizable in infinite volume.
In QCD, if there would appear divergence at the surfaces,
they could be subtracted by the boundary counterterms which
consist of gauge invariant local operators of three or less dimensions.
For fermions, such terms are only the bilinear operators of dimension
three and are given in general as
\begin{eqnarray}
\Delta S_{boundary}
&=&
\int_{0}^{L}d^3y \, \bar\Psi(\y,T) Z_T \Psi(\y,T)
+
\int_{0}^{L}d^3y \, \bar\Psi(\y,0) Z_0 \Psi(\y,0) \ec
\label{staggered:surface:counterterm}
\end{eqnarray}
where $Z_t$ (t=0,$T$) are some matrices.

To determine $Z_0$ and $Z_T$
in the lattice regularization with the staggered fermion,
we should take into account of the symmetry of the staggered fermion
in the finite volume space.
We should consider the following three symmetry transformations:

(i) discrete spatial rotation on
(i,~j)-plain (i,~j=1,~2,~3~)\cite{mitra-wise}
\begin{eqnarray}
x &\to &A^{-1}x \, ; \nonumber\\
&&(A^{-1}x)_i=x_j, (A^{-1}x)_j=-x_i,
\; \mbox{\rm rest unchange,}\\
\psi(x) &\to & S(A)\otimes R^{T}(A) \, \psi(x) \ec\nonumber \\
\bar\psi(x) &\to &\bar\psi(x) \, S(A)^{-1}\otimes [R^{T}(A)]^{-1} \ec
\end{eqnarray}
with
\begin{equation}
S(A)
= \frac{1}{\sqrt{2}}(1-\gamma_i \gamma_j ) \ec   \quad
R^{T}(A)
= \frac{1}{\sqrt{2}}(\gamma_i^\ast-\gamma_j^\ast ) \ee
\end{equation}

(ii) $(\gamma_4 \otimes \gamma_{5}^{\ast})$-parity transformation
\cite{mitra-wise}
\begin{eqnarray}
\psi (\x,x_4)&\to &(\gamma_4 \otimes \gamma_{5}^{\ast}) \,
\psi (-\x,x_4) \ec \nonumber \\
\bar{ \psi }(\x,x_4)&\to &
\bar{\psi} (-\x,x_4) \, (\gamma_4 \otimes\gamma_{5}^{\ast}) \ee
\end{eqnarray}

(iii) $(\gamma_5 \otimes \gamma_5^\ast)$-chiral transformation
(when massless)\cite{kluberg}
\begin{eqnarray}
\psi(x) &\to & e^{i\alpha (\gamma^5 \otimes \gamma^{5\ast})} \,
\psi(x)~,\nonumber \\
\bar\psi(x) &\to & \bar\psi(x) \,
e^{i\alpha (\gamma^5 \otimes \gamma^{5\ast})}
\ee
\end{eqnarray}

\noindent
By these symmetries, the general form of $Z$ is restricted as
\begin{eqnarray}
Z
&=&
c_4 (\gamma_4 \otimes \mbox{\bf 1} )
+c_5 (\gamma_5 \otimes \gamma_4^\ast\gamma_5^\ast)
+\bar c_4 (\gamma_4 \otimes \bar{\gamma}^{\ast}\gamma_4^\ast)
+\bar c_5 (\gamma_5 \otimes \bar{\gamma}^{\ast}\gamma_5^\ast)
\nonumber\\
&=&
c_0 (\gamma_4 \otimes \mbox{\bf 1} ) P_0
+c_1 (\gamma_4 \otimes \mbox{\bf 1} ) P_1
+\bar c_0 (\gamma_5 \otimes \bar{\gamma}^{\ast}\gamma_5^\ast) P_0
+\bar c_1 (\gamma_5 \otimes \bar{\gamma}^{\ast}\gamma_5^\ast) P_1
\ee
\end{eqnarray}
where $\bar\gamma\equiv \sum_{i=1}^{3}\gamma_i $ \
and $c$'s are arbitrary constants.
Then, taking into account of the boundary condition
(\ref{staggered:four-spinor:boundary:inhomogeneous}),
the boundary counterterms can be written as
\begin{equation}
\Delta S_{boundary}= \Delta S^{(0)} + \Delta S^{(T)} \ec
\label{staggered:surface:counterterm:inhomogeneous}
\end{equation}
\begin{eqnarray}
\Delta S^{(0)}
&=& \
\int_{0}^{L}d^3y \, \big\{ \,
c_0^{(0)}
\bar\eta^{ }_0(\y) (\gamma_4 \otimes \mbox{\bf 1} ) P_0 \Psi(\y,+0)
+
c_1^{(0)}
\bar\Psi(\y,+0) (\gamma_4 \otimes \mbox{\bf 1} ) P_1 \eta^{ }_0(\y)
\, \big\}
\nonumber\\
&+&
\int_{0}^{L}d^3y \, \big\{ \,
\bar c_0^{(0)}
\bar\Psi(\y,+0)(\gamma_5 \otimes \bar{\gamma}^{\ast}\gamma_5^\ast)
P_0 \Psi(\y,+0)
+
\bar c_1^{(0)}
\bar\eta^{ }_0(\y) (\gamma_5 \otimes \bar{\gamma}^{\ast}\gamma_5^\ast)
P_1 \eta^{ }_0(\y)
\, \big\}
\ec
\label{staggered:surface:counterterm:inhomogeneous:initial}
\end{eqnarray}
\begin{eqnarray}
\Delta S^{(T)}
&=& \
\int_{0}^{L}d^3y \, \big\{ \,
c_0^{(T)}
\bar\Psi(\y,T-0) (\gamma_4 \otimes \mbox{\bf 1} ) P_0 \eta^{ }_T(\y)
+
c_1^{(T)}
\bar\eta^{ }_T(\y) (\gamma_4 \otimes \mbox{\bf 1} ) P_1 \Psi(\y,T-0)
\, \big\}
\nonumber\\
&+&
\int_{0}^{L}d^3y \, \big\{ \,
\bar c_0^{(T)}
\bar\eta^{ }_T(\y)(\gamma_5 \otimes \bar{\gamma}^{\ast}\gamma_5^\ast)
P_0 \eta^{ }_T(\y)
\nonumber\\
&&\qquad\qquad\qquad\qquad\qquad\quad
+
\bar c_1^{(T)}
\bar\Psi(\y,T-0) (\gamma_5 \otimes \bar{\gamma}^{\ast}\gamma_5^\ast)
P_1 \Psi(\y,T-0)
\, \big\}
\ee
\label{staggered:surface:counterterm:inhomogeneous:final}
\end{eqnarray}
These expressions give the general form of the boundary counterterms
which may appear
in the lattice regularization with the staggered fermion.

According to the above expressions,
we note that even in the case of the
homogeneous Dirichlet boundary condition
there remain the surface counterterms,
which give the coupling between the Neumann components.
\begin{eqnarray}
\Delta S_{surface}
&=& \
\bar c_0^{(0)}
\int_{0}^{L}d^3y \,
\bar\Psi(\y,+0)(\gamma_5 \otimes \bar{\gamma}^{\ast}\gamma_5^\ast)
P_0 \Psi(\y,+0)
\nonumber\\
&+&  \,
\bar c_1^{(T)}
\int_{0}^{L}d^3y \,
\bar\Psi(\y,T-0) (\gamma_5 \otimes \bar{\gamma}^{\ast}\gamma_5^\ast)
P_1 \Psi(\y,T-0)
\ee
\label{staggered:surface:counterterm:continuum}
\end{eqnarray}
The symmetry of the staggered fermion
(in terms of the four-component spinor)
does not seem to be able to exclude these counterterms.
This mean that
the boundary condition
(\ref{staggered:four-spinor:continuum:boundary:homogeneous})
could not be upheld\cite{symanzik} and
the \sch\ of QCD in the lattice regularization
with staggered fermion would not be well-defined.

This is not the case, however.
To examine this problem, we consider the boundary counterterms
at finite lattice spacing.
Noting
$P_0\psi(\y,+0)=\sitarel{\lim}{b\rightarrow0}\frac{1}{b^3}P_0\psi(\x,+1)$
and the similar equation for $P_1 \Psi(\y,T-0)$,
we obtain the lattice counterparts of $\Delta S_{surface}$
as
\begin{eqnarray}
\Delta S_{surface}^\prime
&=& \
\bar c_0^{(0)}
\sum_x \,
\bar\psi(\x,+1)(\gamma_5 \otimes \bar{\gamma}^{\ast}\gamma_5^\ast)
P_0 \psi(\x,+1)
\nonumber\\
&+&
\bar c_1^{(T)}
\sum_x \,
\bar\psi(\x,m-1) (\gamma_5 \otimes \bar{\gamma}^{\ast}\gamma_5^\ast)
P_1 \psi(\x,m-1)
\nonumber\\
&=&
\frac{\bar c_0^{(0)}}{2}
\sum_{\bn}\sum_{k=1}^{3} \, 
\eta_{k}(\bn,2) \,
 \bar{\chi}(\bn,2)
[  \chi(\bn-\hat{\mbox{\boldmath $k$}},2)
 - \chi(\bn+\hat{\mbox{\boldmath $k$}},2) ]
\nonumber \\
&+&
\frac{\bar c_1^{(T)}}{2}
\sum_{\bn}\sum_{k=1}^{3} \, 
\eta_{k}(\bn,2m-1) \,
 \bar{\chi}(\bn,2m-1)
[  \chi(\bn-\hat{\mbox{\boldmath $k$}},2m-1)
 - \chi(\bn+\hat{\mbox{\boldmath $k$}},2m-1)]
\ee
\nonumber\\
\label{staggered:surface:counterterm:lattice}
\end{eqnarray}
{}From this expression, we can see the following
nature of the surface counterterms.
They are local and ${\cal O}(1)$
in terms of the variables $\psi(x)$ and $\bar\psi(x)$ on the
lattice with the spacing~$b$.
On the other hand,
they are derivative terms and ${\cal O}(a) $
in terms of the variables $\chi(n)$ and $\bar\chi(n)$ on the
lattice with the spacing $a$.
As a result,
if the link variables of gauge field were introduced
on the lattice with the spacing $b$, the surface divergence
could occur and we would need the above counterterms.
On the other hand,
as far as they are introduced
on the lattice with the spacing $a$,
these counterterms may not be necessary.

It is worth noting that the similar thing happens to the self-energy
of the staggered fermion.
As shown
in \cite{thun}\cite{mitra-wise}\cite{gockeler},
if the link variables are defined on the finer lattice,
the self-energy correction induced by
the gauge interaction is proportional to the bare mass.
Then the chiral limit is achieved when the bare mass vanishes.
However, if they are defined on the blocked lattice, a
linear-divergent mass term is actually induced
and we need a fine tuning to obtain massless fermion.

Note also that the above result is consistent with the case
in which
the dimensional regularization is adopted for the continuum theory
with the boundary condition
(\ref{staggered:four-spinor:boundary:inhomogeneous}).
In this case,
a kind of parity becomes symmetry of the theory
(including the boundary condition).
It is given by the following transformation,
\begin{eqnarray}
\psi (\x,x_4) &\to&
(\gamma_4 \otimes \gamma_4^{\ast}) \psi (-\x,x_4),
\nonumber \\
\bar{ \psi }(\x,x_4)&\to &
\bar{\psi} (-\x,x_4)(\gamma_4 \otimes\gamma_4^{\ast})
\ee
\end{eqnarray}
By virtue of this symmetry,
the surface counterterms are eliminated.

{}From these considerations, we can expect that
as far as the gauge field is introduced on the
lattice with the spacing $a$,
the boundary counterterms are given in the following form
in the continuum limit.
\begin{eqnarray}
\Delta S_{boundary}
&=&
\int_{0}^{L}d^3y \, \big\{
c_0^{(0)}
\bar\eta^{ }_0(\y) (\gamma_4 \otimes \mbox{\bf 1} ) P_0 \Psi(\y,+0)
+
c_1^{(0)}
\bar\Psi(\y,+0) (\gamma_4 \otimes \mbox{\bf 1} ) P_1 \eta^{ }_0(\y)
\big\}
\nonumber\\
&+&
\int_{0}^{L}d^3y \, \big\{
c_0^{(T)}
\bar\Psi(\y,T-0) (\gamma_4 \otimes \mbox{\bf 1} ) P_0 \eta^{ }_T(\y)
+
c_1^{(T)}
\bar\eta^{ }_T(\y) (\gamma_4 \otimes \mbox{\bf 1} ) P_1 \Psi(\y,T-0)
\big\}
\ee
\nonumber\\
\label{staggered:surface:counterterm:inhomogeneous:a:continuum}
\end{eqnarray}

\subsection*{\normalsize \bf Quark Masses}

For the staggered fermion,
which is equivalent to four-flavored quarks,
the degenerate mass $M$ can be naturally introduced as
\begin{equation}
S_M
= \sum_{x_4=0}^{m} \sum_{\x} M
  \bar{\psi}_a^{\ \alpha}(x) \psi_{\alpha}^{\ a}(x)
=\sum_{n\in \Gamma} \frac{1}{2} M \bar{\chi}(n) \chi (n)
\ee
\label{staggered:free-action:mass}
\end{equation}
Such degenerate mass term for the staggered fermion
does not cause any essential modification of the above
formulation of the \sch.
$S_M$ is simply added to $S_F^{tot}$.

Towards the more realistic QCD, we should lift the mass degeneracy
of four flavor quarks.
This possibility for the staggered fermion has been discussed
by several
authors\cite{mitra}\nocite{mitra:nucl}\nocite{mitra-wise}--\cite{gockeler}.
If we adopt the representation for the gamma matrices such as
$\gamma_4 = \sigma_3 \otimes \mbox{\bf 1}$ and
$\gamma_i = \sigma_1 \otimes \sigma_i$ for $i=1,2,3$,
the non-degenerate diagonal mass terms can be expressed as
\begin{eqnarray}
M_a \delta_{ab}= m  \, \mbox{\bf 1}_{ab}
                + m_4 \, (\gamma_4^\ast)_{ab}
                + i m_{12} \,  (\gamma_2^\ast \gamma_1^\ast)_{ab}
                + i m_{124} \, (\gamma_4^\ast \gamma_2^\ast \gamma_1^\ast)_{ab}
\ec
\end{eqnarray}
In terms of the one-component field $\chi(x)$,
it reads
\begin{eqnarray}
S^\prime_M
&=& \sum
\sum_{\x} M_a
   \bar{\psi}_a^{\ \alpha}(x) \psi_{\alpha}^{\ a}(x)
\\
\nonumber
&=& \sum
\left\{
 \half m \bar{\chi}(n) \chi (n)
+\half m_4 (-)^{n_1+n_2+n_3} \eta_4(n) \bar{\chi}(n) \chi (n+(-)^{n_4} \hat4)
\right.
\\
\nonumber
&&
+i\half m_{12} (-)^{n_1+n_2} \eta_1(n) \eta_2( n+(-)^{n_1} \hat1)
               \bar{\chi}(n) \chi (n+(-)^{n_1} \hat1 +(-)^{n_2} \hat2)
\\
\nonumber
&&
+i\half m_{124} (-)^{n_3}
\eta_4(n) \eta_1( n+(-)^{n_1} \hat1)
          \eta_2( n+(-)^{n_1} \hat1+(-)^{n_2} \hat2)
\nonumber
\\
&&\qquad\qquad\qquad\qquad\qquad\qquad\qquad
\times  \left. \bar{\chi}(n) \chi (n+(-)^{n_1} \hat1+(-)^{n_2} \hat2
                                                     +(-)^{n_4} \hat4)
\right\}
\ee
\label{staggered:free-action:mass-nondegenerate}
\end{eqnarray}
We can see that the general non-degenerate mass terms
lead to the non-local terms extending to the next-to-nearest
neighbors.
They also break the cubic rotational symmetry in the four dimensional
Euclidean lattice.
If we require the spatial cubic symmetry and the locality in time
($n_4$) direction, the non-degeneracy is necessarily reduced
to two\cite{mitra-wise}.
Such mass is given by
\begin{eqnarray}
M_{ab}= m  \, \mbox{\bf 1}_{ab}
        + \frac{1}{\sqrt{3}} m^\prime \, ( \bar\gamma^\ast)_{ab}
\ec
\end{eqnarray}
where $\bar\gamma\equiv \sum_{i=1}^{3}\gamma_i $.
The two eigenvalues of $M_{ab}$ are given by
$M_0=m+m^\prime$ and $m_0=m-m^\prime$.
In terms of the one-component field $\chi(x)$, it looks like
\begin{eqnarray}
S^{\prime\prime}_M
&=& \sum_{x_4=0}^{m} \sum_{\x} M_{ab}
   \bar{\psi}_a^{\ \alpha}(x) \psi_{\alpha}^{\ b}(x)
\\
&=& \sum_{n\in \Gamma}
\left\{
\half m \bar{\chi}(n) \chi (n)
+
\frac{1}{2\sqrt{3}}
m^\prime (-)^{n_1+n_2+n_3+n_4}
\sum_i (-)^{n_i} \eta_i(n)
\bar{\chi}(n) \chi (n+(-)^{n_i} \hat i )
\right\}
\ee
\label{staggered:free-action:mass-nondegenerate:two}
\end{eqnarray}
Note that this mass term, like the degenerate mass
(\ref{staggered:free-action:mass-nondegenerate}),
does not cause any essential modification of the above
formulation of the \sch.
$S_M^{\prime\prime}$ is simply added to $S_F^{tot}$.

The degenerate mass term is local and gauge invariant.
For the non-degenerate case
(\ref{staggered:free-action:mass-nondegenerate:two}),
the gauge invariance of the mass term can be
assured by the suitable insertion of the link variables.
As shown in \cite{thun}\cite{mitra-wise}\cite{gockeler}, if the link
variables are defined on the finer lattice with the lattice
spacing $a$,
the self-energy correction induced by
the gauge interaction is proportional to the bare mass.
Then the chiral limit is achieved when the bare mass vanishes.

\subsection*{\normalsize \bf Discussion}

In the classical continuum limit of the lattice QCD
with the staggered fermion in finite volume,
we obtained the action
(\ref{staggered:schrodinger-func:total-action:four-spinor:continuum:gauge})
supplemented by the boundary condition
(\ref{staggered:four-spinor:continuum:boundary:homogeneous}).
We also determined the possible form of the
boundary counterterms as
(\ref{staggered:surface:counterterm:inhomogeneous:a:continuum})
in the case that
the gauge field is introduced on the lattice with the spacing $a$.
This way of introducing the gauge field
also keeps the chiral property
of the staggered fermion\cite{thun}\cite{mitra-wise}\cite{gockeler}.

In the application to the determination of
$\alpha_s$ in QCD by the finite size scaling technique,
it seems a practical choice to adopt the
homogeneous boundary condition in the numerical
calculation of $\alpha_s$\cite{sint}.
For this boundary condition, the boundary counterterms
(\ref{staggered:surface:counterterm:inhomogeneous:a:continuum})
vanish.
This means that the \sch\ of QCD with the four-flavored quarks
which is supplemented by the homogeneous boundary condition
(\ref{staggered:four-spinor:continuum:boundary:homogeneous})
can become finite by the renormalization of the gauge coupling constant
and the quark mass.
As to the quark mass,
the degenerate mass $M$ will be the first practical choice.
The extrapolation to the massless limit $M\rightarrow 0$
can reduce the number of parameters to be renormalized
in the continuum limit.
Towards the more realistic QCD, we should lift the mass degeneracy
of the four-flavor quarks.
It may be possible to lift the degeneracy
between the two ``light''quarks
and the two ``heavy'' quarks
by using the non-degenerate mass term
(\ref{staggered:free-action:mass-nondegenerate:two}).
Even though,
this is rather crude description of the real quark masses.
This kind of problem of the nonzero quark mass
is yet an open question in the practical numerical
calculation for QCD in finite volume.

There are many works remaining to be done.
Especially, as the next step,
one should calculate
the contribution of the quark loop to the finite part
of the Schr\"odinger Functional, $\Gamma_1[B]$ in
ref. \cite{luscher},
to the first nontrivial order in the weak gauge coupling expansion.
The calculation in two different regularization schemes
would verify the
universality of the Schr\"odinger Functional of QCD.

\begin{acknowledgements}

 We would like to express our sincere thanks to A.~Ukawa and
S.~Aoki for valuable discussions and comments.
We are also indebted to M.~Shimono for discussions and
advice in preparing the manuscript.
\end{acknowledgements}

\end{document}